# Piezo-optomechanical cantilever modulators for VLSI visible photonics

**Mark Dong,**[1,2,6] **David Heim,**[1] **Alex Witte,**[1] **Genevieve Clark,**[1,2] **Andrew J. Leenheer,**[3] **Daniel Dominguez,**[3] **Matthew Zimmermann,**[1] **Y. Henry Wen,**[1] **Gerald Gilbert,**[4,7] **Dirk Englund,**[2,5,8] **and Matt Eichenfield**[3,9]

[1]*The MITRE Corporation, 202 Burlington Road, Bedford, Massachusetts 01730, USA*
[2]*Research Laboratory of Electronics, Massachusetts Institute of Technology, Cambridge, Massachusetts 02139, USA*
[3]*Sandia National Laboratories, P.O. Box 5800 Albuquerque, New Mexico, 87185, USA*
[4]*The MITRE Corporation, 200 Forrestal Road, Princeton, New Jersey 08540, USA*
[5]*Brookhaven National Laboratory, 98 Rochester St, Upton, New York 11973, USA*
[6]*mdong@mitre.org*
[7]*ggilbert@mitre.org*
[8]*englund@mit.edu*
[9]*meichen@sandia.gov*

**Abstract:** Visible-wavelength very large-scale integration (VLSI) photonic circuits have potential to play important roles in quantum information and sensing technologies. The realization of scalable, high-speed, and low-loss photonic mesh circuits depends on reliable and well-engineered visible photonic components. Here we report a low-voltage optical phase shifter based on piezo-actuated mechanical cantilevers, fabricated on a CMOS compatible, 200 mm wafer-based visible photonics platform. We show linear phase and amplitude modulation with 6 $V_\pi$-cm in differential operation, -1.5 dB to -2 dB insertion loss, and up to 40 dB contrast in the 700 nm - 780 nm range. By adjusting selected cantilever parameters, we demonstrate a low-displacement and a high-displacement device, both exhibiting a nearly flat frequency response from DC to a peak mechanical resonance at 23 MHz and 6.8 MHz respectively, which through resonant enhancement of Q~40, further decreases the operating voltage down to 0.15 $V_\pi$-cm.

# 1. Introduction.

There is currently a growing demand for very large-scale integration (VLSI) photonic circuits [1,2] that provide precise, rapid, and low-power control of visible optical fields. Quantum information applications from quantum computing and networks to sensing [3–5] increasingly rely on atom [6–8] and atom-like [9–11] systems, which make use of optical transitions in the visible wavelength regime. In chemical sensing and imaging, visible light is required to interact with particular molecular species [12] and to achieve higher resolution than possible with longer wavelengths. A leading approach for large-scale optical control is programmable Mach-Zehnder meshes (MZMs) [1], built from cascaded Mach-Zehnder interferometers (MZI) (Fig. 1a). Each MZI performs the unitary operation $U(2)$ as the fundamental building block for different types of meshes [13–16], such as multi-port interferometers (Fig. 1b) and binary trees (Fig. 1c). The complexity of scaling these circuits requires high-quality individual MZIs and has led to the development of many modulation schemes. In the near-infrared (NIR), phase modulation in MZIs has been demonstrated at large-scale with thermo-optic phase shifters [17–21] and in individual devices using free-carriers [22], $\chi^{(2)}$ nonlinearities [23–25], and MEMS [26]. In the visible regime, previous reports on thermo-optic [27–29] and thin-film lithium niobate [30] MZIs show promise for VLSI photonics, but there remains an open challenge to build reliable MZMs that satisfy application requirements of high switching bandwidths (>10 MHz), high contrasts (>40 dB), and low losses (<1 dB) per modulator.

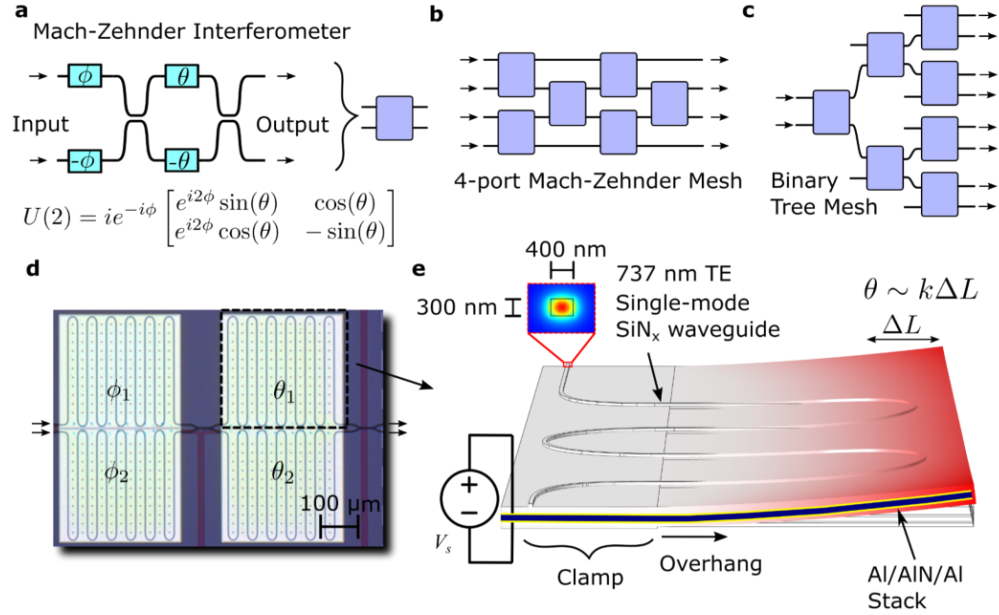

Fig. 1. Piezo-optomechanical cantilevers for large-scale visible photonics a) diagram of a Mach-Zehnder interferometer with four phase shifters in push-pull configuration performing the $U(2)$ operation b) schematic of a 4-port Mach-Zehnder mesh c) schematic of a 2x8 binary tree mesh d) optical microscope image of a fabricated device with four integrated cantilever phase shifters, each controlling the phases $\theta_{1,2}, \phi_{1,2}$ e) operating principles of a piezo-optomechanical cantilever phase shifter, showing that an applied voltage $V_s$ across an aluminum nitride piezo imparts a path-length change to the integrated SiN$_x$ waveguides, inducing an effective phase shift $\theta$

To address this need, we previously introduced a programmable MZM platform [31] based on visible-spectrum silicon nitride (SiN$_x$) waveguides with high-speed (>100 MHz) aluminum nitride (AlN) piezo modulation [32]. However, the modulators in this mesh have a high voltage-loss product (175 V-dB), defined as $\text{VLP} = V_\pi \times \alpha_m$, where $V_\pi$ is the voltage required for a π-phase shift and $\alpha_m$ is the modulator insertion loss. The VLP metric governs the limit on possible mesh circuit depths given maximum voltage (e.g. set by CMOS driver circuitry) and optical loss requirements. Conversely, for a set mesh size, cascading high-VLP modulators to reduce $V_\pi$ or improve unitary fidelity [33] may not be possible due to increasing photon losses. As the number of optical components in a mesh generally increases quadratically with the number of input/output fields [15], modulators with low VLPs are highly desirable.

In this work, we demonstrate a visible-spectrum phase and amplitude modulator using piezo-actuated mechanical cantilevers. An improved undercut process in the fabrication enables reliable, singly clamped cantilevers with large released regions (>500 μm) and lower VLPs in the 20 - 30 V-dB range, an order of magnitude improvement over our previous work [31,34]. The optically broadband modulator has a 6 $V_\pi$-cm, up to >40 dB extinction, low hold-power consumption (<30 nW), -1.5 dB to -2 dB insertion loss, and minimal modulation losses. Moreover, the modulator exhibits a nearly flat frequency response from DC to a peak mechanical eigenmode (up to tens of MHz) for nanosecond switching or resonantly enhanced actuation to further reduce operating voltage (down to 0.15 $V_\pi$-cm or 0.8 V-dB). We arrange the phase modulators for differential operation [13] in an MZI configuration (microscope image shown in Fig. 1d) with the four possible phase shifts labeled. The device consists of 400 nm wide x 300 nm thick SiN$_x$ waveguides coupled to an AlN piezo stack (Fig. 1e). The modulator operates by applying a voltage $V_s$ across the piezo layer, which mechanically deforms the cantilever and induces a path length change $\Delta L$ and phase shift $\theta$ in the waveguides. We characterize the device across the 700 nm - 780 nm wavelength range and explore different cantilever designs to target various operating regimes.

## 2. Phase shifter fabrication and design

We illustrate the cantilever design in Fig. 2. A scanning electron microscope (SEM) image is shown in Fig. 2a of the fully fabricated and released cantilever with false-colored SiN$_x$ waveguides, which are looped several times across the surface of the cantilever to increase the phase shifter response. Fig. 2b maps out a cross-section of our entire layer stack.

The fabrication is based on a 200 mm-wafer optical lithography process at Sandia National Labs which we briefly summarize. First, a bottom aluminum metal layer (M1) is patterned and etched for routing electrical signals and grounds. We then deposit and pattern a sacrificial amorphous-Si (a-Si) layer for defining the cantilevers. Next, a stack of aluminum, aluminum nitride, aluminum (Al/AlN/Al) forms the electrodes and piezo layers for optomechanical actuation. After some buffer oxide, we deposit and etch the SiN$_x$ waveguides to form the optical waveguides. We next etch a set of small release holes through the entire stack (Fig. 2b), exposing additional a-Si to facilitate device release. Finally, post wafer dicing, a xenon difluoride (XeF$_2$) process removes the a-Si, undercutting all cantilever devices on a single die.

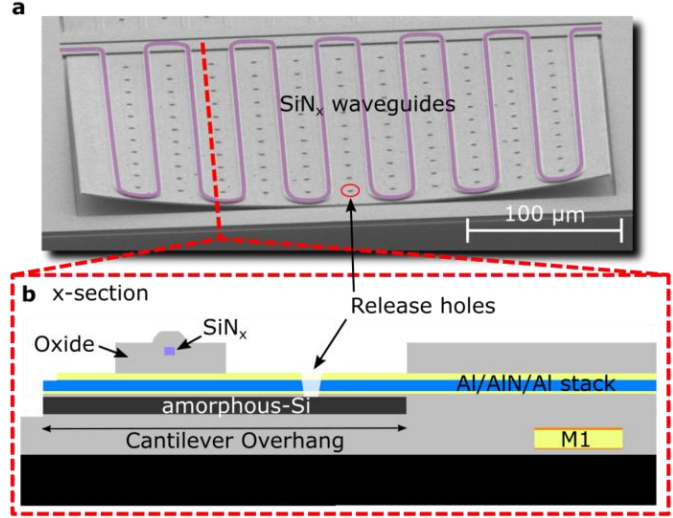

Fig. 2. Basic design and layer stack of a piezo-optomechanical cantilever phase shifter a) SEM of 300 μm overhang length cantilever with SiN$_x$ waveguides colored purple b) cross-section schematic of cantilever, depicting the cantilever overhang defined by the sacrificial amorphous-Si layer, SiN$_x$ waveguide (purple) and oxide cladding (grey), aluminum (yellow) / aluminum nitride (blue) piezo stack, and the aluminum routing metal M1; etched release holes are shown to enable removal of the amorphous-Si layer for large overhangs

The physical mechanisms that contribute to the optical phase shift is primarily due to waveguide path length deformations induced by applying voltages across the piezo layer, in addition to stress-optic effects [35]. Using finite-element models of our cantilever geometry, we calculate the displacement tensor $\nabla \mathbf{u}$, defined as the gradient of the mechanical displacement vector field $\mathbf{u}$, for a given applied voltage $V_s$. Integrating the displacement tensor along the meandering waveguide path (Fig. 2a), which we define as a curve $C$, we find the path length change $\Delta L$ to be

$$\Delta L = \int_C ds \, \hat{s} \cdot \nabla \mathbf{u} \cdot \hat{s} \qquad (1)$$

where $\hat{s}$ is the unit vector parallel to the path $C$. This length deformation then induces a phase shift

$$\theta = 2\pi n_{eff} \Delta L / \lambda \qquad (2)$$

where $n_{eff} = 1.68$ is the effective modal index of our waveguide. Based on our simulations, for an $h=30$ μm overhang cantilever at $V_s = 10$ V, we estimate a total $\Delta L = 0.89$ nm for a single waveguide loop, corresponding to $\theta \sim 0.004\pi$ radians at 737 nm wavelength. We note in the linear elastic regime (applicable for the small strain values present in our system), $\Delta L$ scales approximately linearly with cantilever overhang $h$ and the number of waveguide loops $N_L$.

The induced phase shift's dependence on the cantilever geometry allows for a trade-off between device size, operating voltage, optical losses, and mechanical resonance frequency. We design two different cantilever geometries with characteristics shown in Table 1. Design 1 is a high-displacement cantilever, optimized for DC, low voltage operation with a lower peak mechanical frequency, while design 2 is a low-displacement cantilever, optimized for AC, fast switching with a higher peak mechanical frequency.

**Table 1: Characteristics of Piezo-Optomechanical Cantilevers**

| Device | Overhang $h$ (μm) | Waveguide loops $N_L$ | Peak resonance frequency (MHz) | Footprint (μm$^2$) | Voltage-loss product VLP (V-dB) |
|---|---|---|---|---|---|
| Design 1 cantilever (high-displacement) | 300 | 6 | 6.8 | 350 x 325 | 22 |
| Design 2 cantilever (low-displacement) | 80 | 19 | 23.3 | 100 x 650 | 36 |

## 3. Device Characterization

We characterize our cantilever modulator's performance by measuring MZIs with both design 1 and design 2 parameters by actuating the two internal phase shifters per MZI, each contributing a phase of $\theta_{1,2}$ (Fig. 1c), while the additional phase shifts $\phi_{1,2}$ are unused. We use a 250-μm pitch fiber array to couple a broadly tunable continuous-wave (CW) Ti:Sapphire laser into our SiN$_x$ waveguides through on-chip gratings designed for the 700 nm - 780 nm range. DC and AC electrical signals are delivered with a ground-signal-ground (GSG) RF probe touching down onto electrical pads connected to the phase shifters for active modulation. Insertion losses measured at 737 nm wavelength typically range from -1.5 dB to -2 dB per modulator after subtracting the grating coupler efficiencies.

*3.1 Design 1: DC Actuation*

We first characterize a design 1 MZI by applying a single voltage signal $V_s$ connected in opposite polarities to the two phase shifters such that nominally, $\theta_1 = -\theta_2$. Fig. 3 shows the normalized optical transmission from the MZI's cross-port as the voltage $V_s$ is swept from 0 V - 30 V at 0.25 V step size. We plot modulation performances across 705 nm, 737 nm, and 780 nm wavelengths (Fig. 3a) and find the V$_\pi$ values via a sinusoidal fit of the data to be 14.0 V, 15.2 V, and 16.3 V respectively, increasing with wavelength. The total SiN$_x$ waveguide length for design 1 is 3.9 mm (accounting for all loops), thus we calculate V$_\pi$L ranging from 5.5 V-cm to 6.3 V-cm. The passive directional couplers in our modulator are optimized (50:50 splitting) around 737 nm, thus the depth of modulation decreases as the wavelength moves farther away from this wavelength [33]. The splitting ratios, seen more clearly in log scale (Fig. 3b), vary by wavelength and dip below 40 dB for 780 nm while 737 nm and 705 nm show 28 dB - 30 dB respectively. We attribute the variation to differences in polarization and frequency stability of the laser at different wavelength set points.

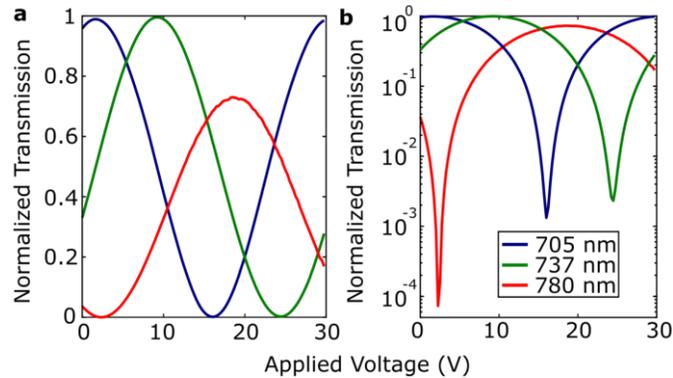

Fig. 3. DC actuation a piezo-optomechanical modulator with design 1 parameters a) normalized transmission of cross-port at 705 nm, 737 nm, and 780 nm b) log scale of same data

*3.2 Design 2: AC Actuation and Mechanical Resonance Enhancement*

We next investigate a design 2 MZI to determine the temporal response and mechanical resonances present in the cantilever. For the experiments in this section, we apply an AC signal to modulate phase shifter $\theta_1$ only, while the other phase shifter $\theta_2$ is set to a specific DC bias point depending on the measurement.

The switching behavior of our modulator is characterized by applying various switch signals. Here, the phase $\theta_2$ is biased such that the modulator turns "on" and "off" as $\theta_1$ is modulated. When a simple square wave is applied (Fig. 4a), we observe many excited mechanical resonances, including a long-lived oscillation at ~23 MHz. The high frequency components in the sharp square edge can be suppressed by tailoring a smoothed (hyperbolic tangent) switch

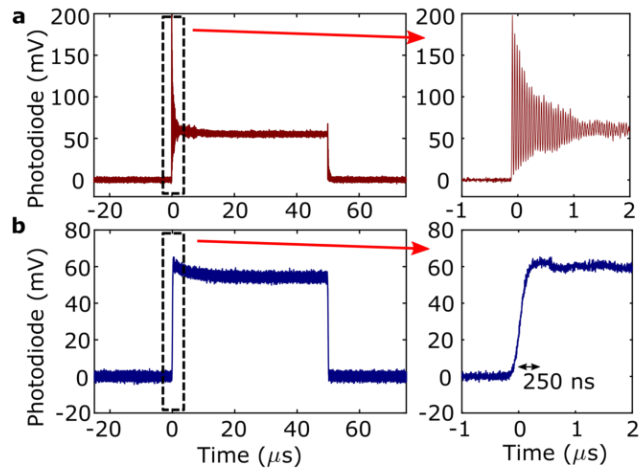

Fig. 4. Switching characteristics of a piezo-optomechanical modulator with design 2 parameters a) cross-port time-trace (16 averages) of an applied 10 kHz square wave showing long-lived mechanical resonances b) cross-port time-trace (16 averages) of an applied 10 kHz tanh square showing a smooth transition with a rise time of 250 ns

signal, resulting in a clean transition with a 250 ns rise time (Fig. 4b) more suitable for applications requiring faster time scales.

We next measure the modulator's frequency transfer function using small-signal (0.5 V pk-pk) sinusoids on $\theta_1$, while setting $\theta_2$ to the maximum slope of the MZI's amplitude response for enhanced contrast. Fig. 5a plots the device's modulation amplitude as the small-signal sine is swept in frequency, normalized to the DC response. Several piezo-mechanical resonances [36] are clearly seen at 1.8 MHz, 4.4 MHz, 8.3 MHz, 14.1 MHz, and the long-lived 23.3 MHz resonance responsible for the oscillations observed in Fig. 4a. Finite-element modeling of the cantilever confirms eigenmodes close to the measured frequencies, showing the resonances belong to the same family of modes. We show two lower order resonances at 1.8 MHz and 4.4 MHz (Fig. 5b), simulated on a cantilever subsection, to illustrate the mechanical deformations. The number of ripples along the free-hanging portion of the cantilever increases for the higher frequency eigenmodes. We note the measured resonance peaks are similar to those observed in other piezo-electronic systems [37,38].

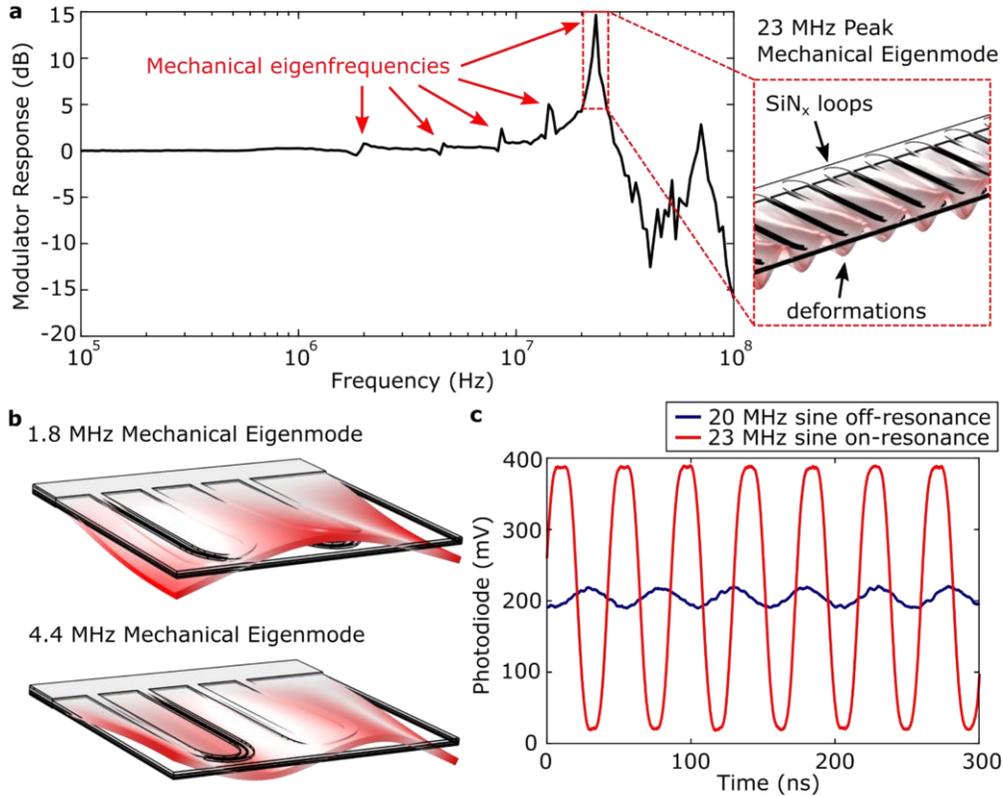

Fig. 5. Frequency response of a piezo-optomechanical modulator with design 2 parameters a) measured small-signal transfer function of the modulator, depicting several mechanical eigenfrequencies including a peak at 23 MHz highlighted with a 3D finite-element simulation of the eigenmode b) additional finite-element simulations of the second and third order mechanical eigenmodes at 1.8 MHz and 4.4 MHz respectively c) time trace of the modulator response driven with a 0.8 V pk-pk sinusoid at 20 MHz (off resonance) and 23 MHz (on resonance), showing a resonantly-enhanced phase shift per volt.

The presence of cantilever mechanical eigenmodes particular to each phase shifter allows for the resonances to greatly enhance the phase shift per voltage response. We focus on the peak mode at 23.3 MHz for which the mechanical ripples and the waveguide loops are spatially aligned approximately in a 1:1 ratio. We record a time-resolved trace of the cross port output while applying sine waves at 20 MHz (off-resonance) and 23 MHz (on-resonance) to $\theta_1$ (Fig. 5c). A large enhancement (~15 dB) of the modulator response is seen due to the mechanical resonance effects. By adjusting the amplitude of the applied sine wave until the modulator output saturates, we measure the single cantilever $V_\pi$ to be 0.8 V. The total $SiN_x$ waveguide length for design 2 is 3.62 mm, corresponding to a $V_\pi L$ of 0.3 V-cm (or 0.15 V-cm for two cantilevers in differential operation). Comparing the resonant $V_\pi$ to the static $V_\pi$ of a single cantilever (36 V) for design 2 (see Supplement), the mechanical Q is estimated to be ~40. The interaction between the optics and the mechanical resonance further contributes to the path displacement effect as well as strain-optic effects – we are currently investigating the detailed theory of the resonant piezo-optoelectronic physics.

### 4. Discussion

We presented two specific designs of our piezo-optomechanical modulator, highlighting its versatility and overall suitability for large programmable photonic mesh circuits in the visible regime. The robustness of our fabrication process enables reliable cantilever performance and engineering of several important device parameters, including peak resonance frequency and $V_\pi$. We characterize additional cantilevers with varying overhang lengths from three different batches of wafers. Fig. 6a, b shows the measured single-loop $V_\pi$ values at DC and peak cantilever resonance, respectively. Each data point is the average of three to five different cantilever modulators, with +/- 1 standard deviation error bars shown. Based on a least-squares fit, both the DC $V_\pi$ and peak mechanical resonance $f_R$ have a predictable inverse relationship with cantilever overhang, given by:

$$V_\pi = a_V / (N_L h) \qquad (3)$$

$$f_R = a_R / h \qquad (4)$$

where $h$ is the cantilever overhang, $N_L$ is the number of waveguide loops, and $a_V$ and $a_R$ are the slope coefficients of the $V_\pi$ and peak resonance equations, respectively. We calculate $a_V$ and $a_R$ to be 42.7 V-mm-loops and 1.81 MHz-mm. From Equations 3 and 4, the critical parameters of $V_\pi$ and $f_R$ are quickly estimated by simply dividing $a$ by the cantilever overhang and in the case of $V_\pi$, further divided by the number of waveguide loops. Unlike $V_\pi$, the peak mechanical resonance of the cantilever does not strongly depend on the number of waveguide loops. This behavior is explained by the resonance mode deformations (Fig. 5b) which is affected predominantly by the density of loops (nominally constant across all measured devices) over the cantilever area. The MHz-range mechanical resonance frequency has an uncertainty of less than one linewidth from the measured devices, making them robust to fabrication variations compared to typical optical resonance structures [39].

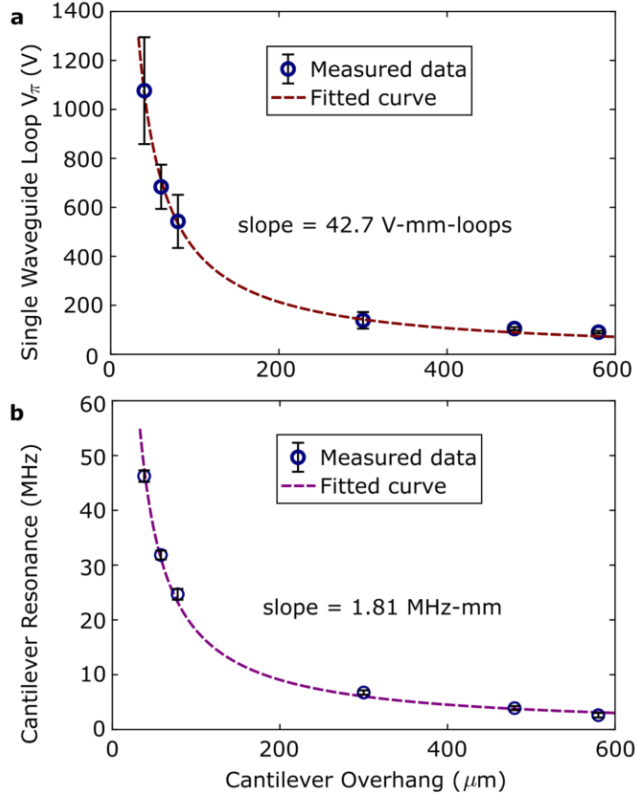

Fig. 6. Dependence of DC $V_\pi$ and peak resonance frequency on cantilever overhang a) measured $V_\pi$ values for various cantilevers with a fitted slope of $a_V$ = 42.7 V-mm-loops b) measured resonance frequency values for various cantilevers with a fitted slope of $a_R$ = 1.81 MHz-mm.

The broad ranges in operating voltage and bandwidth available to our cantilever modulator by simply adjusting parameters $h$ and $N_L$ allow for the engineering of larger photonic meshes to application-specific needs. Ultra-low-$V_\pi$ MZMs are promising candidates for monolithically integrated photonics and CMOS electronic drivers [40] – a single-chip solution which allows for a small number of electronic inputs to control a large number of complex circuits. Optogenetics [41] and display technologies [42] would not require >1 MHz responses and would be well served by a larger cantilever with lower actuation voltage and power consumption. Other applications such as optical switches and optical neural networks [43] would benefit from shorter cantilevers with >10 MHz resonance frequencies for high-speed reconfiguration. Moreover, quantum network switches likely prefer modulators with shorter waveguides and low optical loss at the expense of higher drive voltages. Finally, driving multiple engineered cantilevers on-resonance would be beneficial for phased arrays and light ranging applications [44,45], which require fast and cyclical control of many output beams.

**Funding.** MITRE Quantum Moonshot Project; DARPA ONISQ program; Brookhaven National Laboratory supported by U.S. Department of Energy, Office of Basic Energy Sciences, under Contract No. DE-SC0012704; NSF RAISE TAQS program; Center for Integrated Nanotechnologies, an Office of Science User Facility operated by the U.S. Department of Energy Office of Science.

**Acknowledgments.** M.D. thanks Adrian Menssen, Ian Christen, and Artur Hermans for helpful technical discussions. M.D. also thanks Julia M. Boyle for characterizing the electrical impedances.

**Disclosures.** The authors declare no conflicts of interest.

**Data availability.** Data underlying the results presented in this paper are not publicly available at this time but may be obtained from the corresponding authors upon reasonable request.

**Supplemental document.** See Supplement for supporting content.

# Piezo-optomechanical cantilever modulators for VLSI visible photonics: Supplement

**Mark Dong,**[1,2,6] **David Heim,**[1] **Alex Witte,**[1] **Genevieve Clark,**[1,2] **Andrew J. Leenheer,**[3] **Daniel Dominguez,**[3] **Matthew Zimmermann,**[1] **Y. Henry Wen,**[1] **Gerald Gilbert,**[4,7] **Dirk Englund,**[2,5,8] **and Matt Eichenfield**[3,9]

[1]*The MITRE Corporation, 202 Burlington Road, Bedford, Massachusetts 01730, USA*
[2]*Research Laboratory of Electronics, Massachusetts Institute of Technology, Cambridge, Massachusetts 02139, USA*
[3]*Sandia National Laboratories, P.O. Box 5800 Albuquerque, New Mexico, 87185, USA*
[4]*The MITRE Corporation, 200 Forrestal Road, Princeton, New Jersey 08540, USA*
[5]*Brookhaven National Laboratory, 98 Rochester St, Upton, New York 11973, USA*
[6]*mdong@mitre.org*
[7]*ggilbert@mitre.org*
[8]*englund@mit.edu*
[9]*meichen@sandia.gov*

## S1. Additional Device Performance Data

*S1.1 Power Consumption and Modulation Losses*

We calculate the hold-power consumption for design 1 (the larger of the two designs) based upon measured values of device capacitance $C = 40.5$ pF (from both phase shifters) and on-chip routing metal resistance $R_{chip} = 21\,\Omega$. Previous sheet resistance measurements in the AlN piezo layer show an estimated leakage resistance of 7.5 GΩ at room temperature. The hold-power at $V_\pi$ is then ~30 nW. The energy required for a full switching cycle is $E = CV_\pi^2 = 9$ nJ, or 9 mW electrical power consumption at a 1 MHz switching rate.

We observe negligible modulation losses, or amplitude-phase coupling when actuating the phase shifters. Fig. S1 plots the measured power in μW of bar, cross, and the sum of both ports. The total output varies by an upper bound of ~1% for the ~one minute duration of the DC scan, which also includes laser frequency and polarization instabilities, grating coupler mismatch of the bar and cross ports, and drifts of the fiber arrays.

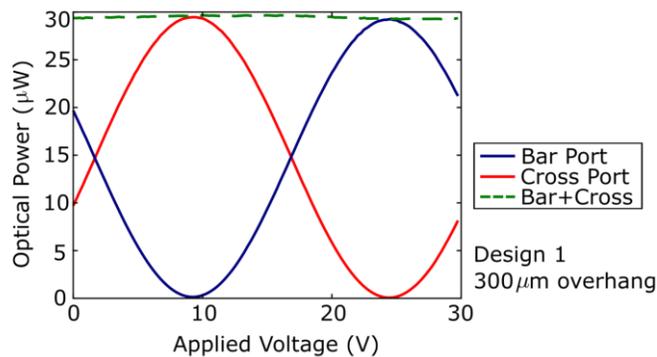

Fig. S1. DC actuation a cantilever modulator with design 1 parameters, plotting the measured power transmitted through the bar port, cross, port, and sum of both ports.

*S1.2 Additional Measurements for Design 1 and Design 2 Cantilever Modulators*

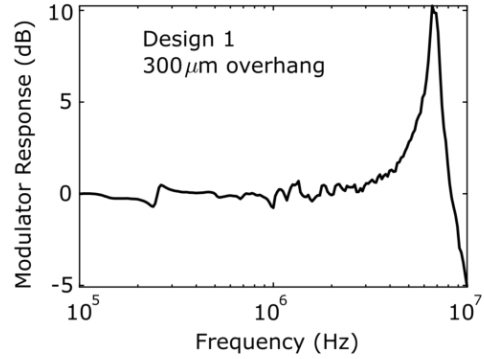

Fig. S2. Frequency response of a cantilever modulator with design 1 parameters. There is a strong peak resonance frequency at 6.8 MHz.

To complete the characterization of designs 1 and 2 presented in the main text, we show additional AC or DC measurements of each design. Fig. S2 plots the frequency response of a design 1 cantilever, illustrating a similar shape to that of design 2 with lower resonance frequencies. Here, the peak resonance frequency is at 6.8 MHz for design 1.

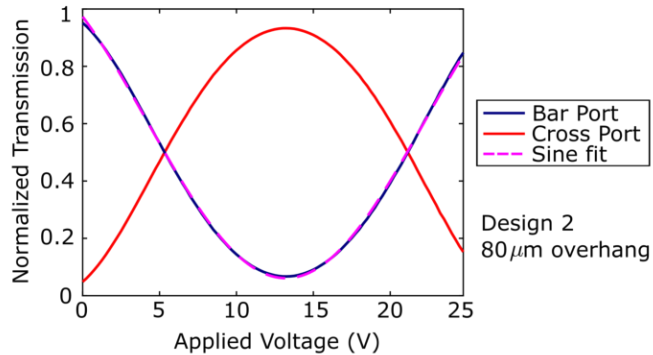

Fig. S3. DC actuation a cantilever modulator with design 2 parameters. A fitted sine wave shows the linearity of the phase shifter, with a $V_\pi$ of 18 V.

Fig. S3 plots the bar and cross ports a push-pull DC voltage scan (same scan as in section 3.1 in the main text) for a design 2 cantilever. The $V_\pi$ is higher due to the shorter overhang, but compensated by the larger number of waveguide loops. A sine fit to the bar port highlights the linearity of the phase shifter with respect to the driving voltage. Adjusting for the length of the waveguide (3.62 mm), we measure a 6.5 $V_\pi$-cm for this device, similar to that of design 1.